# Translation and discussion of the De Iride, a treatise on optics by Robert Grosseteste

**Amelia Carolina Sparavigna**
Department of Applied Science and Technology, Politecnico di Torino, Italy

*Here I am proposing a translation and discussion of the De Iride, one of the short scientific treatise written by Robert Grosseteste. In the first part of his Latin text we find reflection and refraction of light, described in a geometrical optics. In the second part, Grosseteste is discussing the rainbow and how the colors are created.*

Robert Grosseteste was an English scientist and philosopher of the Middle Age. He was born into an humble Anglo-Norman family in the county of Suffolk in England. He was Bishop of Lincoln from 1235 AD till his death, on 9 October 1253. Considered one of the most prominent and remarkable figures in thirteenth-century, he was a man of many talents. As told in [1], he was commentator and translator of Aristotle and Greek thinkers, philosopher, theologian, and student of nature. About physics, Grosseteste wrote several short works. Besides his studies, as a bishop, he focused his energies on rooting out abuses of the pastoral care. He is considered one of the three Oxonians that played a relevant role in the revival of Optics in the thirteen century [2]. After him there were Roger Bacon and John Peckham, who considered Grosseteste as an inspiration for their scientific developments.

Grosseteste is also considered as a thinker that played a key role in the development of scientific method. In [1] it is reported that A.C. Crombie [3] claimed Grosseteste as the first in the Latin West to develop an account of an experimental method in science, and that he made a systematic use of the method of "experimental verification and falsification". Moreover, Crombie remarked that Grosseteste gave a special importance to mathematics in explaining the physical phenomena. These claims however have been the subject of considerable debate.

In Ref.1, it is told that the Grosseteste's experimental method was quite different from a method of controlled experiment. Grosseteste made no use of such a method in his writings, deriving his conclusions on the basis of a mix of considerations, appealing to authority and everyday observation (the Latin "experimentum"). He made use of thought experiments and certain metaphysical assumptions, such as the assumption of a principle of "least action", that we will find here in reading the De Iride, one of his scientific treatises. Grosseteste used the empirical observation as one factor for his discussion of nature. However, he is far from employing an experimental method involving a controlled experiment: we can assume that his experimental "verification and falsification" was as a first step towards the modern method.

As it is told in [1], reporting the studies of Ludwig Bauer [4], Grosseteste gave a relevant role to mathematics in attempting to explain the physical world. In his treatise On Lines, Angles and Figures, Grosseteste remarks that "the consideration of lines, angles and figures is of the greatest utility since it is impossible for natural philosophy to be known without them …. All causes of natural effects have to be given through lines, angles and figures, for otherwise it is impossible for the reason why, the propter quid, to be known in them" [1,4]. In the treatise, On the Nature of Places, a continuation of the treatise On Lines, Angles and Figures, Grosseteste remarks that "the diligent investigator of natural phenomena can give the causes of all natural effects, therefore, in this way by the rules and roots and foundations given from the power of geometry".

Undoubtedly, Grosseteste saw a key role for geometry in the explanation of natural phenomena. As remarked in [3], Grosseteste was deeply concerned with a detailed investigation of natural phenomena: it was his attitude of mind, and his emphasis on the importance of mathematics, that was a stimulus to thinkers in the Oxford of the fourteenth-century, who were developing the beginnings of a mathematical physics.

In a recent paper [5], I have shortly discussed the role of the light in the creation of the world as seen by Grosseteste. Here I am translating and discussing one of the works of Grosseteste on optics, entitled De Iride, On Rainbow. In fact he is not only discussing the rainbow. In the first part of the text, he discusses reflection, refraction and optical instruments. In the second part he is proposing the rainbow as a phenomenon of refraction of light. He explains how the shape of the rainbow is originated and the formation of the colors.

Here, I am subdividing the Latin text in several sections [6]. For each section, it is reported the original text and it is given translation, where who is writing, ACS, applied her knowledge of Latin. Some additional comments are given too.

The Latin text is given in MS UI Gothic characters.

1. INC: Et perspectivi et physici est speculatio de iride. EXPL: Et similiter secundum alias connumerationes claritatis et obscuritatis luminis et puritatis et impuritatis diaphani satis manifestae sunt secundum colores omnes arcus varii variationes.

*INC: Optics and Physics are speculating on the rainbow.*
*EXPL: And likewise, are reasoning about other facts on brightness and obscurity of the light of purity and impurity on transparent media, and all we know about the bows of various colors according to the variations of the medium.*

I have translated "perspectivus" as "optical", like in Ref.2. In [7], it is told that Perspective, in the sense of the "science of optics," came in English from Old French perspective and directly from Medieval Latin perspectiva ars "science of optics," from fem. of perspectivus "of sight, optical" from Latin perspectus "clearly perceived," pp. of perspicere "inspect, look through, look closely at," from per- "through" + specere "look at". In the sense of "art of drawing objects so as to give appearance of distance or depth" is first found 1590s, influenced by Italian prospettiva, an artists' term. The figurative meaning "mental outlook over time" is first recorded 1762. The "iris" is a flowering plant, also "prismatic rock crystal," from L. iris (pl. irides) "iris of the eye, iris plant, rainbow," from Greek iris (gen. iridos) "a rainbow; the lily; iris of the eye," originally "messenger of the gods," personified as the rainbow. The eye region was so called (early 15c. in English) for being the colored part.

2. Et perspectivi et physici est speculatio de iride. Sed ipsum "quid" physici est scire, "propter quid" vero perspectivi. Propter hoc Aristoteles in libro meteorologicorum non manifestavit "propter quid", quod est perspectivi, sed ipsum "quid", de iride, quod est physici, in brevem sermonem coarctavit. Ideoque in praesenti ipsum "propter quid", quod attinet ad perspectivum, pro modulo nostro et temporis opportunitate suscepimus explicandum.

*It is of optics and physics to speculate about the rainbow. But, the same "what" the physics needs to know, is a "because of what" the optics needs. And in fact, Aristotle, in the book on the meteorology, did not show "because of what", in the sense of optics, but "what" is the rainbow, which is physics, in a quite short discussion. Hence in this paper, this "because of what", concerning optics, is started discussing and explaining, in our manner and time opportunity.*

Here we have "quid" (interrogative pronoun [8]), "what", that is the effect, or the phenomenon, the physics needs to describe. The "propter quid", "because of what", is instead an answer given by the research, on the causes of the phenomenon. In the Latin text, we have also "modulo nostro". Modulus is a "small measure," dim. of modus "measure, manner".

3. In primis igitur dicimus, quod Perspectiva est scientia, quae erigitur super figuras visuales, et haec subalternat sibi scientiam, quae erigitur super figuras, quas continent lineae et superficies radiosae, sive proiecta sint illa radiosa ex sole, sive ex stellis, sive ex aliquo corpore radiante. Nec putandum, quod egressio radiorum visualium sit positio imaginata solum absque re, sicut putant illi, qui partem considerant et non totum. Sed sciendum, quod species visibilis est substantia assimilata naturae solis lucens et radians, cuius radiatio coniuncta radiationi corporis lucentis exterius totaliter visum complet.

*First then, let us say that optics is a science, which is based on the figures of the visual perceptions, and it is subaltern to the science, which is based upon figures and schemes, which contain lines and radiating surfaces, being them cast by the radiating sun, or by stars, or by any other radiant body. And it has not to be thought that the going out of visual rays from eyes is only a virtual argument, without any reality, as people who consider "the part and not the whole" are arguing. But let us note that visible objects are of a nature similar to the nature of the shining and sparkling sun, the radiation of which, combined with the radiation of the external surface of a body, completes the total perspective of vision.*

First of all [7], the noun "figure", is the "visible form or appearance of a person," from Old French figure (10 century) "shape, body, form, figure, symbol, allegory," from Latin figura "a shape, form, figure". Originally in English with meaning "numeral," but sense of "form, likeness" is almost as old (mid-13 century). And "species", that from 1550s, is a classification in logic, here is meaning "kind, sort," originally "appearance, sight, a seeing," related to specere "to look at, to see, behold". Therefore I translated as "object".

4. Unde philosophi naturales tangentes id, quod est ex parte visus naturale et passivum, dicunt visum fieri intussuscipiendo. Mathematici vero et physici considerantes ea, quae sunt supra naturam, tangentes id, quod est ex parte visus supra naturam et activum, dicunt visum fieri extramittendo. Haec partem visus, quae fit per extramissionem, exprimit Aristoteles aperte in libro de animalibus ultimo dicens: "oculus profundus videt remote; nam motus eius non dividitur, neque consumitur, sed exit ab eo virtus visualis et vadit recte ad res visas." Et iterum in eodem: "Tres dicti sensus scilicet visus, auditus, olfactus, exeunt ab instrumentis, sicut aqua exit a canalibus, et propter hoc longiores nasus sunt boni olfactus."

*Therefore, some philosophers handling this natural things, are considering the natural visual perception as passive, that is, as an "intro-mission". However, mathematicians and physicists, concerning the nature of visual perception, consider that it occurs according an "out-emission". Now, this part of the sight, which is effected by an out-emission, Aristotle plainly discussed in the last chapter of his book on the animals, that "the back of the eye sees far away; from its emission it is not divided, nor consumed, but its ability of sight goes forward from him and right to the things we are seeing." And again, in the same: "Three are our conscious*

*senses, namely, sight, hearing, smell, they come out from the organs, just as it emerges from the water in canals, and therefore a long nose has a good smelling."*

5. Perspectiva igitur veridica est in positione radiorum egredientium.

*In optics, then, the true position concerning the rays is that of their emission.*

Position (n.) [7], as a term in logic and philosophy, is coming in English from the Old French posicion, from Latin positionem (nom. positio) "act or fact of placing, position, affirmation" from posit-, pp. stem of ponere "put, place". Meaning "manner in which a body is arranged or posed" first recorded 1703. Meaning "official station, employment" is from 1890.
We have that Grosseteste used "extramissionem" in section 4 and here "egredientium". So I have softened "out-emission" in "emission". It seems that Grosseteste agreed with the theory of out-emission, but in any case, I suppose that he believde simply in the emission of light from some sources.
About the visual perception, there were two ancient Greek schools, providing a different explanation of vision. The first was proposing an "emission theory": vision occurs by means of rays emanated from the eyes and received by objects. We can see an object directly, or by means of refracted rays, which came out of the eyes, traversed a transparent medium and after refraction, arrive to the object. Among the others, Euclid and Ptolemy followed this theory. The second school proposed the "intro-mission" approach which sees vision as coming from something entering the eyes representative of the object. Aristotle and Galen followed this theory, which seems to have some contact with modern theories [9].
It seems that Grosseteste had mixed Aristotle's ideas with the out-emission theory, and therefore I used simply "emission".

6. Cuius partes principales sunt tres secundum triplicem modum transitionis radiorum ad rem visam. Aut enim transitus radii ad rem visam est rectus per medium diaphani unius generis interpositi inter videntem et rem visam. Aut transitus eius est secundum rectum ad corpus habens naturam huius modi spiritualis, per quam ipsum est speculum, et ab ipso reflectitur ad rem visam. Aut transitus radii est per plura diaphana diversorum generum, in quorum contiguitate frangitur radius visualis et facit angulum, et pervenit radius ad rem visam non per incessum rectum, sed per viam plurium linearum rectarum angulariter conjunctarum.

*Of which (optics), there are three main parts, according to the three ways of transition the rays have to the objects of vision. Either the path of the rays to the visible object is straight through a transparent medium having a specific feature, interposed between who is looking and the object. Or, it is ruled by a path directed to a body having a virtual nature, that is, a mirror, reflected by it, back to the object we are seeing. Or it is the passage of the rays through more transparent media of different kinds, where, at the interfaces, the ray is broken and makes an angle, and the ray comes to the object not with a straight path, but by means of several straight lines, having a number of angles at the related interfaces.*

Transition means the passage from one place to another. Grosseteste is subdividing the propagations of rays in three cases, the first is the direct propagation, the second is the reflection on mirrors and the third the refraction. I rendered "spiritualis" using "virtual".

7. Primam partem "couplet" scientia nominata de visu; secundam illa, quae vocatur de speculis. Tertia pars apud nos intacta et incognita usque ad tempus hoc permansit. Scimus tamen, quod Aristoteles tertiam partem complevit, quae plus ceteris partibus sui subtilitate multo difficilior et naturarum profunditate longe mirabilior extitit. Haec namque pars perspectivae perfecte cognita ostendit nobis modum, quo res longissime distantes faciamus apparere propinquissime positas et quo res magnas propinquas faciamus apparere brevissimas et quo res longe positas parvas faciamus apparere quantum volumus magnas, ita ut possibile sit nobis ex incredibili distantia litteras minimas legere, aut arenam, aut granum, aut gramina, aut quaevis minuta numerare. Qualiter autem haec admiranda contingunt, sic fiet manifestum. Radius visualis penetrans per plura diaphana diversarum naturarum in illorum contiguitate frangitur et eius partes in diversis diaphanis existentes in illorum contiguitate angulariter coniunguntur. Hoc autem manifestum est per experimentum illud, quod ponitur principium in libro de speculis: si in vas mittatur quid, sumatur distantia, ut iam non videatur et infundatur aqua, videbitur, quod immissum est. Manifestatur etiam illud idem per hoc, quod subiectum continui est corpus unius naturae; radium igitur visualem in contiguitate duorum diaphanorum diversi generis necesse est a contiguitate decidere. Cum autem totalis radius a principio uno sit generatus, nec possit penitus continuitas illius solvi, nisi interrupta esset eius generatio, necesse est, ut in contiguitate duorum diaphanorum non sit completa radii discontinuatio; medium autem inter plenam continuitatem et completam discontinuitatem non potest esse nisi punctus unius contingens duas partes non directe, sed angulariter.

*The first part of this science is named "de visu", the second "about mirrors". The third part is coming in our possession unknown and untouched. We know, however, that Aristotle had discussed this third part, which is the much more difficult, and the subtlety of which was by far the more remarkable, emerging from the depths of the nature. This part of optics, if fully understood, shows us the way in which we can made objects at very long distance appear at very close distance, and large things, closely situated, appear very small, and small things at a certain distance we can see as large as we want, so that, it is possible for us to read the smallest letters at incredible distance, or count the sand, or grain, or grass, or anything else so minute. In what way, however, it is necessary to understand how this wonderful happens, so it will become clear to everybody. Visual rays penetrating through several transparent different materials, are broken at interfaces; and the parts of these rays, in the different existing transparent materials, at the interface of those are angularly connected. This, however, is clear by means of an experience, the principle of it is set down in the book on the mirrors: if we cast an object into a vessel, and the distance is assumed that it may not be seen, and some water poured into, it will be seen what is inside. The same is displayed by a body having a continuous nature too; therefore, the visual ray, at the interface of two transparent media with different features, is subjected to a contiguity law. When one total ray is generated from a source, the continuity of it cannot be broken, unless its generation is broken, and at the interface of two transparent media, the ray is not discontinuous; at the interface, we cannot have a full continuity and a complete discontinuity and therefore, at each point of the interface the two parts are, not directly, but angularly connected.*

Couplet (n.) [7], from the Latin copula "tie, connection". I supposed that Grosseteste was telling that the first part of optics is coupled with the direct propagation of rays.
In this part of the treatise we find the description of some phenomena that we can obtain with lenses; he seems to describe, for instance, a magnifying glass useful to see the small things or read the small letters in a book. And then I am supposing that Grosseteste had some lenses in his "laboratory". Moreover, he tells that "we can made things at very long distance appear at very close distance, and large things closely situated appear very small, and small things at a certain we can see as large as we want". Had he a sort of telescope?

In any case, we can suppose that he had some reading stones. A reading stone was a more or less hemispherical lens, that was placed on a text to magnify the letters, so that people with presbyopia could read. Reading stones were among the earliest common uses of lenses. According to Wikipedia, [10] they were developed in the 8th century, by Abbas Ibn Firnas. The function of reading stones was replaced by the use of spectacles from the late 13th century onwards. Early reading stones were made from rock crystal (quartz) as well as glass.

The earliest written records of lenses date to Ancient Greece. In his play, The Clouds (424 BCE), Aristophanes is mentioning a burning-glass, a lens used to focus the sun's rays to produce fire. Pliny the Elder show that burning-glasses were known to Romans, [11] and mentions what was probably a corrective lens: Nero was said to watch the gladiatorial games using an emerald, probably concave to correct for myopia [12]. Pliny is also describing the magnifying effect of a glass globe filled with water. And here too, Grosseteste is describing a globe filled with water. What is interesting in the Grosseteste description is that he find the reason of these effects in the refractions of the rays.

8. Quanta autem sit radii angulariter adiuncti a recto incessu declinatio, sic imaginabimus. Intelligamus radium ab oculo per medium aeris secundum diaphanum incidentem in continuum et directum protrahi et a puncto, in quo incidit super diaphanum, lineam protrahi in profunditatem illius diaphani, quae cum superficie diaphani ex omni parte faciat angulos aequales. Dico igitur, quod incessus radii in secundo diaphano est secundum viam lineae dividentis per aequalia angulum, quem continet radius imaginabiliter in continuum et directum protractus et linea a puncto incidentiae radii ad angulos aequos super superficiem secundi diaphani in profunditatem eius ducta.

*But how large is the angular deviation from the straight path associated to a ray? Let us consider the ray from the eye through the air medium, incident on a second transparent medium, as a straight line to the point, where it is incident on the transparent medium; then let us make the line deep in the transparent medium, line that makes equal angles with the surface of transparent medium, that is, normal to the interface. I say, therefore, that the prolongation of the ray in the second transparent medium is following a line, separating of a certain angle, which is one half of the angle i obtained as follow. i is the angle given by the line which is the prolongation of the ray, without interruption and direct, drawn away from the point of incidence deep in the medium, equal to the angle i, drawn above the surface of the second transparent medium.*

Here we find the Grosseteste's refraction law. Grosseteste's law is telling that the angle of refraction is one-half the angle of incidence *i*. Of course, it is quite different from the Snell's law that we use, containing the trigonometric functions of angles and the refractive index.

Refraction was studied by the Greek science too. Ptolemy had found a relationship regarding the angles of refraction [13]. Ptolemy found in fact an empirical law, fitting figures with experimental data. He measure the refraction from air to water, and water to glass [14]. Ptolemy plotted *r*, the refractive angle, against *i*, the incident angle, at ten-degree intervals from *i*=0° to *i*=80°. The resulting values of *r* were in agreement with the sine-law. Alhazen, in his Book of Optics (1021), studied the refraction too. Refraction was accurately described by Ibn Sahl, of

Baghdad, in the manuscript On Burning Mirrors and Lenses (984) [13]. He made use of it to work out the shapes of lenses that focus light with no geometric aberrations [13]. The law was rediscovered by Thomas Harriot in 1602, who did not publish his results although. In 1621, Willebrord Snellius (Snell) derived a mathematically equivalent form, that remained unpublished, during his life. René Descartes independently derived the law in terms of sines in 1637, in his treatise "Discourse on Method". After Descartes' solution, Pierre de Fermat proposed the same solution based on his principle of least time.

9. Quod autem sic determinetur anguli quantitas in fractione radii, ostendunt nobis experimenta similia illis, quibus cognovimus, quod refractio radii super speculum fit in angulo aequali angulo incidentiae. Et idem manifestavit nobis hoc principium philosophiae naturalis, scilicet quod "omnis operatio naturae est modo finitissimo, ordinatissimo, brevissimo et optimo, quo ei possibile est".

*So we have determined the amount of the refractive angle of the rays. We know that there are similar experiments giving the refraction of the rays on mirrors, fitting an angle equal to the angle of incidence. And the same tells us that principle of the philosophy of nature, namely, that "every action of the nature is well established, most ordinate, in the best and shortest manner, as it is possible."*

Here we have Grosseteste's principle of "least action". I have translated "finitissimo" with "well established", as given by [8]. The English finite (adj.) is coming from L. finitus, pp. of finire "to limit, set bounds, end," from finis (see finish). But, in Latin, finitus has also the meaning of established, defined, determined [8]. In my opinion, this second meaning was that used by Grosseteste.
It is interesting to note that the Grosseteste's principle is given after a sentence on the reflection of rays from mirrors, that he named refraction. It was in the 17th century, that Pierre de Fermat postulated that "light travels between two given points along the path of shortest time," which is known as the principle of least time or Fermat's principle [15].

10. Res autem, quae videtur per medium plurium perspicuorum, non apparet esse ut ipsa est secundum veritatem, sed apparet esse in concursu radii egredientis ab oculo in continuum et directum protractum et lineae ductae a re visa cadentis in superficiem secundi perspicui propinquiorem oculo ad angulos aequales undique. Hoc autem nobis manifestum est per idem experimentum et consimiles ratiocinationes, quibus novimus, quod res visae in speculis apparent in concursu visus directe protracti et lineae ductae super speculi superficiem ad angulos undique aequales.

*Moreover, the object which is seen through a medium composed of several transparent materials, does not appear to be as it truly is, but it is appearing composed by the concurrence of the rays from the eye, continuous and direct, and by the lines starting from the viewed object and falling on the (second) surface, that is nearest to the eye, according to its normal (the line having equal angles from all the sides). This is clear to us from experiments and similar reasoning that we know, that an object seen in a mirror appears in the concurrence of the propagation of the lines of sight and the lines drawn directly upon the surface of the mirror, normal to this surface.*

Here we can suppose that he had a method to create the images of objects reflected from mirrors and for objects passing through a transparent medium. In the last sentence, he is telling that we can create the image of an object reflected from a mirror, using the rays and the normal to the mirror, as we are used to do in geometric optics.

It is remarkable that Grosseteste does not use in the De Iride a term such as "diopter" or "dioptron" (instrument to look through), which is coming from the Greek. From the Guglielmo Gemoll's dictionary, 1959, we have that διοπτευω, means to observe, consider all sides, explor); διοπτηρ, is the ranger; διοπτρον, the instrument to look through. The ancient dioptra were astronomical and surveying instrument, dating from the 3rd century BCE. The dioptra were a sighting tube or, alternatively, a rod with a sight at both ends, attached to a stand. So, the ancient dioptra usually had not lenses. However, in Italian, we use "diottro", to define the interface between two different optical media. And "diottrica" is the science concerning the light refracted by diaphanous media. In English, the term diopter arrived from French, having the same meaning it has in Italian. Probably Grosseteste knew that the Greek term dioptra was used for surveying; the second sense, that of optical medium, was not yet arrived from French.

11. His itaque manifestis, scilicet quantitate anguli, secundum quem frangitur radius in contiguitate duorum diaphanorum, et loco apparentiae rei visae per medium diaphanorum plurium, adiunctis his principiis, quae sumit perspectivus a philosopho naturali, scilicet quod secundum quantitatem anguli, sub quo videtur aliquid, et situm et ordinem radiorum apparet quantitas et situs et ordo rei visae, et quod magna distantia non facit rem invisibilem, nisi per accidens, sed parvitas anguli, sub quo videtur: patens est perfecte in rationibus geometralibus posito diaphano notae magnitudinis et figurae et notae ab oculo distantiae, qualiter apparebit res notae distantiae et notae magnitudinis et situs secundum locum et magnitudinem et situm; et patens est eisdem modus figurandi diaphana ita, ut illa diaphana recipiant radios egredientes ab oculo secundum quantitatem anguli, quem voluerint, in oculo facti, et restringant radios receptos, quomodocunque voluerint, super res visibiles, sive fuerint illae res visibiles magnae sive parvae, sive longae sive prope positae; et ita appareant eis omnes res visibiles in situ, quo voluerint, et in quantitate, qua voluerint; et res maximas, cum voluerint, faciant apparere brevissimas, et e contrario brevissimas et longe distantes faciant apparere magnas et optime visu perceptibiles.

*It is evident, namely, the quantity of the angle according to which the ray is broken at the interface (contiguity) of the two transparent media, and where the image of an object appears arising from several transparent media; and let us add those principles of optics, which are given by the philosophers studying the natural phenomena, that is, that given the amount of the angle, under which an object is seen, it appears its position and size, according to the order and organization of the rays; and that it is not the great distance rendering a thing invisible, except by accident, but the smallness of the angle under which it is seen: it is clear that it is possible, using geometrical ratios, knowing the position and the distance of the transparent medium, and knowing the distance from the eye, to tell how an object appears, that is, given its distance and size, to know the position and the size of the image; and it is also clear, how to design the shape of the transparent medium, in order that this medium is able to receive the rays coming out from the eye, according to the angle we choose, collected in the eye, and focusing the rays as we like over the observed objects, whether they are large or small, or everywhere they are, at long or short distances; in such a way, all objects are visible, in the position and of the size given by the device; and large objects can appear short as we want, and those very short and at a far distance, on the other hand, appear quite large and very perceptible.*

This is a quite interesting part of the treatise. Here we find that Grosseteste is proposing the geometrical optics, and applied to rays of light, we can give the position and magnitude of the images of objects. Moreover, he is telling that we can obtain some recipes to design the surface of lenses, and arrange some lenses to have a telescope. Again, we can ask ourselves, whether he had actually a telescope or he simply was arguing on its possibility, after studying the descriptions of optical devices in some Arabic manuscripts.

> 12. Et huic tertiae parti perspectivae subalternata est scientia de iride. Non enim possibile est iridem fieri radiis solaribus per incessum rectum a sole in concavitatem nubis incidentibus. Facerent enim in nube illuminationem continuam non secundum figuram arcualem, sed secundum figuram aperturae ex parte solis, per quam ingrederentur radii in nubis concavitatem. Nec possibile est, ut iris fiat per reflexionem radiorum solis super convexitatem rorationis a nube descendentis, sicut super speculum convexum, ita, ut concavitas nubis recipiat radios reflexos et sic appareat iris, quia, si sic esset, non esset iris omnino arcualis figurae, et accideret, quod quanto sol esset altior, tanto iris esset maior et altior, et quanto sol esset dimissior, esset etiam iris minor; cuius contrarium sensui est manifestum. Necesse est ergo, quod iris fiat per fractionem radiorum solis in roratione nubis convexae. Dico ergo, quod exterius nubis est convexum et interius illius est concavum. Quod patet per naturam levis et ponderosi. Et illud, quod apparet nobis de nube, necessario est minor semisphaera, licet appareat in visu semisphaera et cum a concavitate nubis descendat roratio, necesse est illam rorationem in summo esse convexam pyramidaliter, ad terram descendentem, ideoque in propinquitate terrae plus quam in superiori parte condensatam.

*And in the third part of optics we have the study of the rainbow. Undoubtedly, it is not possible the rainbow is given by a direct crossing of the solar rays in the cavities of the clouds. Because the continuous illumination of the cloud does not produce an arc-like image, but some openings towards the sun, through which the rays enter the cavity of the cloud. And it is not possible that the rainbow is produced by a reflection of the rays of the sun upon the surfaces of the raindrops falling down from the cloud, as reflected by a convex mirror, so that the cavity of the cloud receives in this manner the reflected rays, because, if it would be so, the rainbow would not be an arc-like object; moreover, it would happen that increasing the altitude of the sun, the rainbow would be greater and higher, and decreasing the sun altitude, the rainbow would be smaller; this is contrary to what is shown by the experience. It is therefore necessary that the rainbow is created by the refraction of the sun's rays by the humidity of the cloud. Let me tell, therefore, that outside the cloud is vaulted, and inside it is hollow. This is clear from the nature of "light matter" and "heavy matter". And that, what we see of a cloud is smaller than a hemisphere, even though it appears to us as a hemisphere, and when the humidity comes down from inside of the cloud, it is necessary that it assumes the volume of a convex pyramid at the top, descending to the ground, and therefore it is condensed in the proximity of the earth, more than in its upper part.*

Convex [7] in English is coming from French convexe, from Latin convexus "vaulted, arched," pp. of convehere "to bring together". Possibly, it is coming from the idea of vaults carried together to meet at the point of a roof. "Convex lens" is from 1822. Concavity [7], in English from Old French concavité "hollow, concavity", or directly from Latin concavitatem (nom.

concavitas), from Latin concavus "hollow". I have therefore considered the concavity of the cloud, as its hollow parts. The convex part as its arched part.
Roratione in Latin in the drew drop falling. I translated with raindrops and humidity in the air.
For a discussion on the Grosseteste's and Medieval theories on rainbow, see [16].

13. Erunt igitur in universo quattuor diaphana, per quae penetrat radius solis scilicet aer purus continens nubem, secundo nubes ipsa, tertio supremum et rarius rorationis a nube venientis, quarto inferius et densius eiusdem rorationis. Necesse est igitur per ea, quae praedicta sunt de fractione radii et quantitate anguli fractionis in contiguitate duorum diaphanorum, radios solares primo frangi in contiguitate aeris et nubis et deinde in contiguitate nubis et rorationis, ut per has fracturas concurrant radii in densitate rorationis, ibique iterum fracti sicut a cono pyramidali se diffundant non in pyramidem secundi rotundam, sed in figuram assimilatam curvae superficiei pyramidis rotundae expansam in oppositum solis. Ideoque est eius figura arcualis, et apud nos apparet iris australis; et quia conus praedictae figurae est prope terram et ipsius expansio est in oppositum solis, necesse est, ut medietas illius figurae vel amplius cadat in superficiem terrae et reliqua medietas vel minus cadat ex opposito solis in nubem. Ideoque sole existente prope ortum vel occasum apparet iris semicircularis et est maior; sole vero existente in aliis sitibus apparet iris portio semicirculi. Et quanto sol altior, tanto portio iridis minor. Et propter hoc in locis multae accessionis solis ad zenith capitum non apparet omnino iris in hora meridiana. Quod Aristoteles dicit arcum varium apud ortum et occasum solis parvae esse mensurae, non intelligendum est de parvitate quantitatis, sed de parvitate luminositatis, quae accidit propter transitum radiorum per multitudinem vaporum in hac hora plus, quam in horis ceteris. Quod ipse Aristoteles consequenter innuit dicens: hoc esse propter diminutionem eius, quod resplendet de radio solis in nubibus.

*Then, in the universe there are four transparent media, through which the rays of the sun penetrate, that is, pure air containing the cloud, second the cloud itself, third the highest and most rarefied humidity coming from the cloud, and fourth, the lower and denser part of that humidity. From all the things discussed before on refraction and related angles at the interface between two media, it is necessary the rays of the sun are first refracted at the boundary of air and cloud, and then at the boundary of cloud and humidity, so that, after these refractions, the rays are conveyed in the bulk of humidity, and after, they are broken again though its pyramidal cone, however, not assuming the shape of a rounded pyramid, but in the form similar to the curved surface of a rounded pyramid, expanded opposite to the sun. Therefore its shape is that of a bow, and to us (in England), the rainbow never appears in the South, and, because the aforesaid cone is close to the earth, and it is expanding opposite the sun, it is necessary that more than a half of that cone falls on the surface of the earth, and the rest of it falls on the cloud, opposite the sun. Accordingly, on sunrise or sunset, a semicircular rainbow appears and is larger; when the sun is in other positions, the rainbow appears as a portion of the semicircle. And increasing the altitude of the sun, the portion of the rainbow decreases. And for this reason, in those places where the sun can reach the zenith, the rainbow never appears at noon. Aristotle tells that the "quantity" of the different arcs we can see on sunrise and sunset is small, but, Aristotle's small "quantity" is to be understood not concerning the "size" but the luminosity, which happens because the rays are passing, during these hours, through a large quantity of vapor, much larger than in other hours of the day. Aristotle himself suggests as a*

*consequence, that there is a reduction of that which shines because of the rays of the sun in the clouds.*

Here Grosseteste continues his discussion on the rainbow phenomenon. Let us note that Grosseteste uses the term "zenith", which is coming from Arabic. "Et propter hoc in locis multae accessionis solis ad zenith capitum non apparet omnino iris in hora meridiana". Zenith (n.): Reference 7 is telling that it is used in English from the late 14 century, from Middle Latin, cenit, senit, as a bungled scribal transliterations of Arabic samt "road, path," abbreviation of samt ar-ras, lit. "the way over the head." Letter -m- misread as -ni-. The Medieval Latin word could as well be influenced by a rough agreement of the Arabic term with classical Latin semita "sidetrack, side path".

14. Cum autem color sit lumen admixtum cum diaphano, diaphanum vero diversificetur, secundum puritatem et impuritatem, lumen autem quadrifarie dividatur, secundum claritatem scilicet et obscuritatem et tunc secundum multitudinem et paucitatem, et secundum harum sex differentiarum connumerationes sint omnium colorum generationes et diversitates, varietas coloris in diversis partibus unius et eiusdem iridis maxime accidit propter multitudinem et paucitatem radiorum solis. Ubi enim est maior radiorum multiplicatio, apparet color magis clarus et luminosus; ubi vero minor est radiorum multiplicatio, apparet color magis attinens hyacinthino et obscuro. Et quia luminum multiplicatio et a multiplicatione ordinata diminutio non sit, nisi per resplendentiam luminosi super speculum, vel a diaphano, quod per figuram suam in loco quodam congregat lumen et in loco conveniente disgregando diminuit, et haec dispositio receptionis luminis non est dispositio fixa, manifestum est, quod non est in potestate pictorum assimilare iridem, cum tamen sit possibilis eius assimilatio secundum dispositionem non fixam.

*However, the color is light mixed with a transparent medium; the medium is diversified according to the purity and impurity, but the light is fourfold divided; it is to be divided according to the brightness, and of course, to the obscurity, and according to intensity (richness) and tenuity (thinness), and according to the six different enumerations the variety of all the colors is generated, the variety of colors that appear in the different parts of a single rainbow, is mainly due to the intensity or tenuity of the rays of sun. Where there is a greater intensity of light, it appears that the colors are more luminous and bright: but where there is less intensity of light, it appears that the color turns to the dark color of Hyacinthus. And because the intensity of light and the decrease of intensity is not subjected to a rule, except in the case of light shining on a mirror, or passing through a transparent medium, which, by means of its own shape, can gathers the light in a certain place, and, in a certain place can disrupt the light, diminishing it, and the arrangement of receiving the light is not a fixed one, it is clear that that it is not in the skill of an artist to reproduce the rainbow, but it is possible to imitate accordingly to a certain arrangement.*

It seems to me that Grosseteste is telling that we can have convergent or divergent lenses. Or that different images can observed, with respect to the focal planes. And therefore an artist can reproduce the effects created by a mirror, or convergent and divergent lenses; but, for the rainbow, this is too much difficult.

15. Diversitas vero unius iridis ad aliam in coloribus suis tum accidit ex puritate et impuritate diaphani recipientis, tum ex claritate et obscuritate luminis imprimentis. Si enim fuerit diaphanum purum et lumen clarum, erit color eius plus assimilatus albedini et luci. Si vero fuerit diaphanum recipiens habens permixtionem vaporum fumosorum et claritas luminis fuerit pauca, sicut accidit prope ortum et occasum, erit color minoris splendoris et magis obfuscatus. Et similiter secundum alias connumerationes claritatis et obscuritatis luminis et puritatis et impuritatis diaphani satis manifestae sunt secundum colores omnes arcus varii variationes. Explicit tractatus de iride secundum Lincolniensem.

*On the other hand, the difference of the colors of a rainbow from those of other rainbows is due to the purity and impurity of the transparent medium supporting it, as well as from the brightness and obscurity of the light impressing it. If we have a pure transparent medium and bright light, the color is whitish. If the recipient medium is a mixture of vapors and mist and the light is hazy, as occurs near the East and West, the colors are less splendid and their brightness reduced. In the same manner, according to the enumeration of brightness and obscurity of light and of purity and impurity of the medium, all the arcs of various colors can be seen.*
*Here is the end of the discussion on the rainbow, according to a Lincolnian.*

**References**
[1] Lewis, Neil, Robert Grosseteste, The Stanford Encyclopedia of Philosophy (Winter 2010 Edition), http://plato.stanford.edu/entries/grosseteste/
[2] G. ten Doesschate, Oxford and the revival of Optics in the thirteenth century, Vision Rev. Vol.1, 1962, 313-342
[3] Crombie, A. C. (1955). "Grosseteste's Position in the History of Science," in Robert Grosseteste: Scholar and Bishop, ed. Daniel A. Callus, Oxford: Clarendon Press, 98–120.
[4] Ludwig Baur, ed. (1912). Die Philosophischen Werke des Robert Grosseteste, Bischofs von Lincoln, Beiträge zur Geschichte der Philosophie des Mittelalters, 9, Münster: Aschendorff Verlag. [The standard edition of Grosseteste's philosophical works and scientific opuscula]
[5] A.C. Sparavigna, From Rome to the Antipodes: the medieval form of the world, arXiv, 2012, http://arxiv.org/abs/1211.3004
[6] The Latin text is that given y "The Electronic Grosseteste", http://www.grosseteste.com/cgi-bin/textdisplay.cgi?text=de-iride.xml, DE IRIDE, which is reporting the printed Source: Die Philosophischen Werke des Robert Grosseteste, Bischofs von Lincoln, W. Aschendorff, 1912, pp. 72 - 78.
[7] D. Harper, Online Etymology Dictionary, 2012 http://www.etymonline.com/index.php
[8] L. Castiglioni, S. Mariotti, Vocabolario della Lingua Latina, Loescher Ed.,1972.
[9] http://en.wikipedia.org/wiki/Visual_perception
[10] http://en.wikipedia.org/wiki/Reading_stone,
[11] Pliny the Elder, The Natural History, translated by John Bostock, XXXVII, Chap. 10.
[12] Pliny the Elder, The Natural History, translated by John Bostock, XXXVII, Chap. 16
[13] http://en.wikipedia.org/wiki/Snell's_law
[14] A. Mark Smith, Ptolemy and the Foundations of Ancient Mathematical Optics: A Source Based Guided Study, Volume 89, Part 3, American Philosophical Society, 1999
[15] http://en.wikipedia.org/wiki/Principle_of_least_action
[16] Raymond L. Lee, Jr., Raymond L. Lee Alistair B. Fraser, The Rainbow Bridge: Rainbows in Art, Myth, and Science, Penn State Press, 2001